\documentclass[aps,prl,twocolumn,showpacs,superscriptaddress,floatfix]{revtex4-1}
\usepackage{amsmath,amssymb}
\usepackage{graphicx}
\usepackage{epsfig}
\usepackage{color}
\usepackage{rotating}
\usepackage{bm}
\usepackage{setspace}
\usepackage{hyperref}
\begin{document}
\title{Visualizing the elongated vortices in $\gamma$-Ga nanostrips}
\author{Hui-Min Zhang}
\affiliation{State Key Laboratory of Low-Dimensional Quantum Physics, Department of Physics, Tsinghua University, Beijing 100084, China}
\affiliation{Institute of Physics, Chinese Academy of Sciences, Beijing 100190, China}
\author{Zi-Xiang Li}
\affiliation{Institute for Advanced Study, Tsinghua University, Beijing 100084, China}
\author{Jun-Ping Peng}
\affiliation{Institute of Physics, Chinese Academy of Sciences, Beijing 100190, China}
\author{Can-Li Song}
\email[]{clsong07@mail.tsinghua.edu.cn}
\affiliation{State Key Laboratory of Low-Dimensional Quantum Physics, Department of Physics, Tsinghua University, Beijing 100084, China}
\affiliation{Collaborative Innovation Center of Quantum Matter, Beijing 100084, China}
\author{Jia-Qi Guan}
\author{Zhi Li}
\affiliation{Institute of Physics, Chinese Academy of Sciences, Beijing 100190, China}
\author{Lili Wang}
\author{Ke He}
\author{Shuai-Hua Ji}
\author{Xi Chen}
\affiliation{State Key Laboratory of Low-Dimensional Quantum Physics, Department of Physics, Tsinghua University, Beijing 100084, China}
\affiliation{Collaborative Innovation Center of Quantum Matter, Beijing 100084, China}
\author{Hong Yao}
\email[]{yaohong@tsinghua.edu.cn}
\affiliation{Institute for Advanced Study, Tsinghua University, Beijing 100084, China}
\affiliation{Collaborative Innovation Center of Quantum Matter, Beijing 100084, China}
\author{Xu-Cun Ma}
\email[]{xucunma@mail.tsinghua.edu.cn}
\affiliation{State Key Laboratory of Low-Dimensional Quantum Physics, Department of Physics, Tsinghua University, Beijing 100084, China}
\affiliation{Institute of Physics, Chinese Academy of Sciences, Beijing 100190, China}
\affiliation{Collaborative Innovation Center of Quantum Matter, Beijing 100084, China}
\author{Qi-Kun Xue}
\affiliation{State Key Laboratory of Low-Dimensional Quantum Physics, Department of Physics, Tsinghua University, Beijing 100084, China}
\affiliation{Collaborative Innovation Center of Quantum Matter, Beijing 100084, China}
\date{\today}

\begin{abstract}
We study the magnetic response of superconducting $\gamma$-Ga via low temperature scanning tunneling microscopy and spectroscopy. The magnetic vortex cores rely substantially on the Ga geometry, and exhibit an unexpectedly-large axial elongation with aspect ratio up to 40 in rectangular Ga nano-strips (width $l$ $<$ 100 nm). This is in stark contrast with the isotropic circular vortex core in a larger round-shaped Ga island. We suggest that the unusual elongated vortices in Ga nanostrips originate from geometric confinement effect probably via the strong repulsive interaction between the vortices and Meissner screening currents at the sample edge. Our finding provides novel conceptual insights into the geometrical confinement effect on magnetic vortices and forms the basis for the technological applications of superconductors.
\end{abstract}
\maketitle
\begin{spacing}{0.995}
Magnetic vortices in mesoscopic superconductors, including symmetric disks, triangles, squares and rectangles \cite{Baelus2002dependence, Geurts2006symmetric, Zhang2012unconventional, teniers2003nucleation}, have recently stimulated tremendous interest due to their fundamental importance in controlling superconducting energy dissipation, a significant feature for superconductor-based nanotechnologies. When a superconductor is shrunk to nanoscale, exotic vortex shapes (i.e.\ giant vortex and antivortex-vortex molecule) and configurations may occur, but so far mostly inferred from indirect measurements \cite{cordoba2013magnetic}. Direct real space visualization of vortices in nanostructured superconductors remains challenging and rarely explored. A few scanning tunneling microscopy (STM) studies present the preliminary evidence of giant vortex in Pb islands \cite{Cren2009ultimate, Cren2011vortex, Tominaga2013trapping}, but the ill-defined geometry of the samples investigated renders the issue quite complicated and difficult for a direct comparison with theory.

Among all these superconducting nanostructures, rectangular nanostrips or nanowires are especially appealing for superconducting electrical circuit and quantum interference device. Regular oscillations of resistance and critical current as a function of magnetic field have been respectively reported in superconducting InO and Al strips \cite{Johansson2005nanowire, Morgan2015measurement}, which can be accounted for by the so-called Weber blockade theory \cite{Pekker2011webber}. Here each oscillation in either magnetoresistance or critical current corresponds to adding a single vortex into the strip, in analogy with single electron transport through quantum dots in the Coulomb blockade regime. Furthermore, the penetrating vortices are anticipated to configure into a one-dimensional chain at the center of strip \cite{Morgan2015measurement, Pekker2011webber, Lotero2007critical}, because of the lower vortex potential energy there. Upon increasing field, the vortex chain may split into two or more buckled parallel rows \cite{Bronson2006Equilibrium, Palacios1998vortex, Kramer2010direct}. However, so far as we know, such vortex chains have little been directly visualized, and whether the model applies for the extremely narrow superconducting nanostrips remains unjustified.

Herein we report the direct visualization of unexpectedly elongated vortex cores in superconducting $\gamma$-Ga nanostrips by STM, and show that they are caused by the strong repulsive interaction between the vortices and Meissner screening currents at the sample edge. All experiments were conducted in a Unisoku ultrahigh vacuum STM system equipped with molecular beam epitaxy (MBE) for \textit{in-situ} sample preparation. The MBE growth of superconducting Ga nanoislands has been described in detail elsewhere \cite{Zhang2015molecular}, and in the Supplemental Material \cite{supplementary}. The measurements were carried out at 4.4 K unless otherwise noted. A magnetic field up to 7 T can be applied perpendicular to the sample surface. The differential conductance \textit{dI/dV} spectra and maps were measured by disrupting the feedback circuit at the setpoint of $V_{\mathrm{s}}$ = 10 mV and $I$ = 100 pA, sweeping the sample voltage $V_\textrm{s}$, and extracting the conductance using a standard lock-in technique with a bias modulation of 0.3 mV at 987.5 Hz.
\end{spacing}

\begin{figure}[h]
\includegraphics[width=\columnwidth]{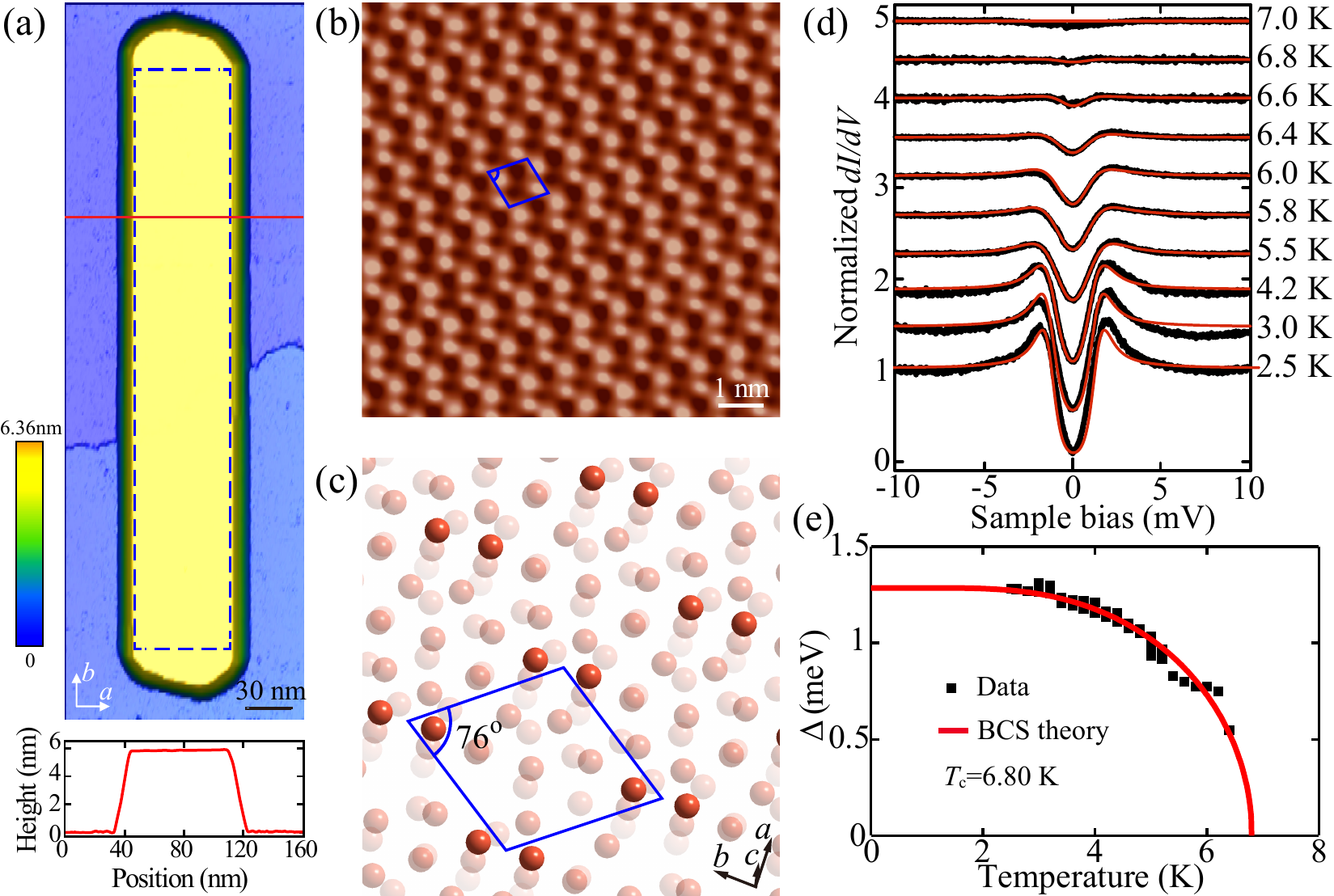}
\caption{(color online) (a) STM topography of a rectangular Ga nanostrip with a height of 5.85 nm (45 ML) ($V_{\mathrm{s}}$ = 3.5 V, $I$ = 50 pA, 483 nm $\times$ 161 nm). Line profile taken along the red curve characterizes the height and width $w$ of Ga nanostrip. Both labels \textit{a} and \textit{b} denote respectively the orientations of the short and long side of the rectangular strips throughout this paper. Dotted rectangle marks the region where all ZBC maps are taken in Fig.\ 2(a). (b) Atomic resolution image ($V_{\mathrm{s}}$ = 20 mV, $I$ = 50 pA, 10 nm $\times$ 10 nm) of the Ga nanostrip. The bright spots correspond to the Ga atoms in the top layer. (c) Schematic crystal structure of $\gamma$-Ga. Along the \textit{c}(001) axis, each unit cell consists of four atomic layers. For clarity, the spheres in each layer are coded in different opacity. The brightest Ga spheres at the top layer form a dimmer-like structure, consistent with the STM image in (b). (d) Temperature-dependent \textit{dI/dV} spectra on Ga nanostrip (black dots) and their best fits to the Dynes's expression (red curves), with the broadening parameter $\Gamma$ of 0.1 $\sim$ 0.2 meV. (e) The extracted gap magnitude $\Delta$ versus temperature (black squares). The red line shows the fit to BCS gap function.
}
\end{figure}

Figure 1(a) shows the constant-current topographic image of a rectangular Ga nanostrip, which has a length $l\sim$ 440 nm and a width  $w\sim$ 64 nm. Note that the nanostrip exhibits an almost perfect rectangle, which can easily be modeled in theory. The atomically resolved STM image on the nanostrip, shown in Fig.\ 1(b), reveals defect-free surface. The unit cell, marked by the blue rhombus, consists of two bright spots with a periodicity of $a_0$ = 0.86 nm and an intersection angle of 76$^\textrm{o}$, which are consistent with $\gamma$-Ga(001) plane [Fig.\ 1(c)] \cite{bosio1973structure}. Figure 1(d) plots a series of differential conductance \textit{dI/dV} spectra acquired on the Ga nanostrip at various temperatures, normalized to the normal-state one above $T_c$ (10 K). A superconducting gap with clear coherence peaks is visible at 2.5 K. At elevated temperature, the gap gradually reduces and both the coherence peaks are suppressed, as expected. The superconducting gaps can be fitted by using Dynes's expression \cite{Dynes1978direct}, as illustrated in Fig.\ 1(d). Figure 1(e) plots the extracted superconducting energy gap $\Delta$ at various temperatures. Fitting the data to BCS gap function yields $T_c$ = 6.80 K and $\Delta(0)$ = 1.28 meV. 

\begin{figure*}[t]
\includegraphics[width=2\columnwidth]{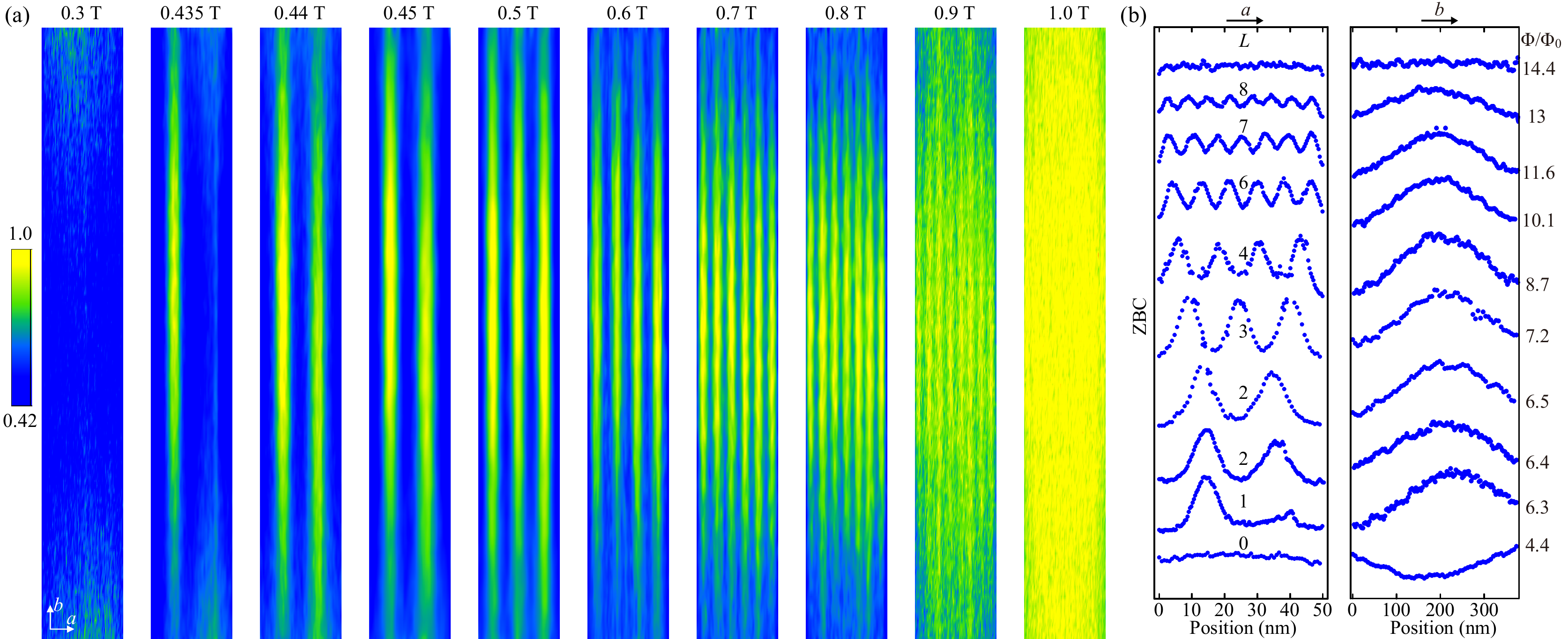}
\caption{(color online) (a) Normalized ZBC maps at various magnetic fields, acquired in a field of view of 380 nm $\times$ 51 nm on the Ga nanostrip of Fig.\ 1(a). The normalization was performed by dividing these maps by the normal-state ZBC value at the vortex core. Yellow regions with enhanced ZBC correspond to the vortex cores, all showing enormous elongation along the \textit{b} axis. The small ZBC enhancement on the right side Ga nanostrip at 0.435 T probably indicates that we are probing a critical state justly before the penetration for the second vortex. (b) ZBC profiles across vortex centers in (a). The ZBC presents oscillations along the \textit{a} axis, while it shows a single extremum along the \textit{b} axis. Here $\Phi/\Phi_0$ and \textit{L} denote the applied magnetic flux and total vorticity in the strip, respectively.
}
\end{figure*}

\begin{figure*}[t]
\includegraphics[width=2\columnwidth]{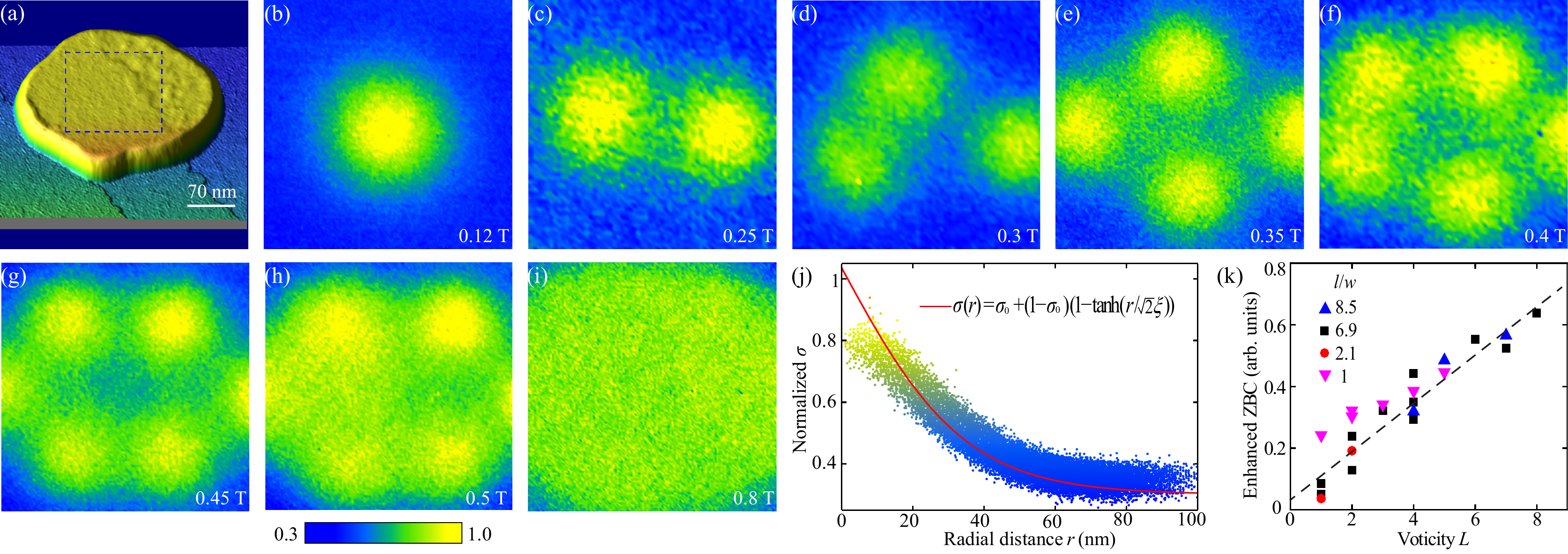}
\caption{(color online) (a) STM topography ($V_{\mathrm{s}}$ = 3.0 V, $I$ = 50 pA) of a larger Ga island (350 nm $\times$ 280 nm) with a height of 3.39 nm ( 26 ML). The dotted square marks the region over which all ZBC maps in (b-i) are acquired. (b-i) Magnetic Field dependence of ZBC maps, acquired in a field of view of 140 nm $\times$ 140 nm in (a). The isotropic vortex cores are unchanged respective of the fields. (b) ZBC map (140 nm $\times$ 140 nm) taken on a round shaped Ga island with a height of 3.39 nm (26 ML), showing a highly isotropic vortex core. (j) Radial dependence of $\sigma(r)$ across the isotropic vortex, which are color-coded to match ZBC map in (b). Here $\sigma(r)$ has been normalized to optimize their fit to the inserted expression by minimizing the residual, as shown by the red curve. (k) Magnetic vortex-induced ZBC enhancement on various Ga nanostructures.
}
\end{figure*}

Having identified the lattice structure and superconductivity, we then focus on the magnetic response of $\gamma$-Ga nanostrip. To visualize the magnetic vortices in real space, we map the spatial variation of zero bias conductance (ZBC) on the dotted rectangle in Fig.\ 1(a), a common approach for STM imaging of vortices \cite{Cren2009ultimate, Cren2011vortex, Tominaga2013trapping, Hayashi1996star, Bergeal2006scanning, ning2009observation, nishimori2004first, Hanaguri2012scanning, zhou2013visualizing, song2011direct, Song2012suppression}. This technique takes advantage of the ZBC contrast within and outside the magnetic vortex cores, and has a higher spatial resolution ($\sim \xi$). Here, the suppressed superconductivity within the vortex cores leads to a higher ZBC value, leaving superconductivity only outside the vortex cores with lower ZBC. Thus the regions with enhanced ZBC indicate the emergence of magnetic vortices. Figure 2(a) illustrates such maps at various fields ranging from 0.3 T to 1 T. Two intriguing phenomena are immediately noticed. The first phenomenon, which is also the more interesting and unexpected one, is that the magnetic vortex cores expand and compress along the longer (\textit{a} axis) and shorter (\textit{b} axis) sides of the rectangular Ga nanostrip, respectively, regardless of the field. This consequently leads to the anisotropic vortex cores with great axial elongation. The elongation ratio $\eta$ is estimated to be 15 at 0.435 T, and increases further with the magnetic field. At 1.0 T, many vortices merge together and the superconductivity is almost completely killed, pushing the whole island into the normal state. Figure 2(b) reveals the field-dependent ZBC profiles across the vortex cores. Along the \textit{b} axis, the existence of only one extremum in every ZBC profile excludes reasonably the possibility of vortex chain for a single individual yellow stripe. On the other hand, ZBC oscillates along the \textit{a} axis, with each peaked region representing single or multiple (e.g.\ giant vortex) of the magnetic flux quantum $\Phi_0$. The number of peaks invariably appears smaller than, but tends to close the applied flux $\Phi/\Phi_0$ as the field increases. Above 0.7 T ($\Phi/\Phi_0$ = 10.1), the ratio between them becomes greater than 0.5, ruling out the possibility of giant vortex. We therefore suggest that each peak (or yellow stripe) actually denotes a single flux quantum $\Phi_0$, namely the vorticity \textit{L} = 1 [Fig.\ S1].

Secondly, we find that the magnetic vortex first penetrates into the Ga nanostrip in a rather high filed, i.e.\ 0.435 T in Fig.\ 2(a). This is caused by the finite energy barrier for vortex entry, consisting of the geometric \cite{Brandt1999geometric} and Bean-Livingston (BL) surface barrier \cite{Bean1964surface}. Both barriers conspire to preclude the penetration of vortices at low field. The first vortex penetration occurs only when the surface superconducting currents exceed the pair-breaking current, locally weakening superconductivity and then allowing the nucleation of a vortex.

The anisotropic internal vortex structure constitutes the major finding in this study: their large elongation and the mechanism behind are unprecedented. For anisotropic vortex cores, such as sixfold star-shape \cite{Hayashi1996star}, fourfold \cite{nishimori2004first, Hanaguri2012scanning, zhou2013visualizing} and twofold symmetry \cite{song2011direct, Song2012suppression, du2014anisotropic}, to name a few, seen previously in certain unconventional superconductors, the anisotropy in either the Fermi surface or the superconducting gap has been suggested to be the primary cause. However, this is not the case here: in $\gamma$-Ga the superconducting gap can be well fitted by the isotropic BCS theory [Fig.\ 1(e)], and no gap anisotropy is involved. To shed more insight into the possible mechanism behind, we have investigated the vortices in more Ga nanostructures, and find that the vortices invariably elongate once if the width of Ga nanostrips is reduced to below $\sim$ 100 nm. In contrast, usual isotropic circular vortices develop in a roughly round-shaped Ga island with a larger lateral dimension [Fig.\ 3(a)], which although shows the same lattice structure and superconductivity as Ga nanostrips. Shown in Figs.\ 3(b-i) are the isotropic vortex cores, which bear strong similarities with those observed in conventional superconductors such as Pb \cite{ning2009observation}. This completely excludes the anisotropy in intrinsic electronic structure of $\gamma$-Ga as a possible cause for the elongated vortices. Instead, the vortex elongates mainly via a geometrical confinement effect.

Figure 3(j) plots the radial dependence of vortex-induced ZBC [or $\sigma(r)$]. Based on the Ginzburg-Landau expression for the superconducting order parameter $\Delta$ near the interface between a superconductor and a normal metal, the radial ZBC profile across the vortex core should obey \cite{Bergeal2006scanning}
\begin{equation} \label{eq:GLvortex}
\sigma(r)=\sigma_{\infty}+(1-\sigma_{\infty})(1-\textrm{tanh}(-r/\sqrt{2}\xi))\,,
\end{equation}
where $\sigma(r)$ is the normalized ZBC away from the vortex core and \textit{r} is the radial distance from the vortex center. A best fit of ZBC profile $\sigma(r)$ to Eq.\ (1) yields a superconducting coherence length $\xi$  = 24.6 $\pm$ 1.1 nm at 4.4 K. Figure 3(k) further qualifies the vortex-induced ZBC enhancement as a function of the vorticity \textit{L} in the round Ga island and rectangular Ga nanostrips, seen more visually in Fig.\ S2. The linear relationship between them, guided by the dashed line, indicates the equivalence (except for the shape) among all observed magnetic vortices, regardless of the geometry.

We now consider possible explanations for the elongated vortices in the Ga nanostrips. Firstly, as the width $w$ of Ga nanostrips is comparable to their Fermi wave length $\lambda_\textrm{F}$, quantum confinement will become important and might lead to sizable spatial modulation in $\Delta$ across the width of nanostrips\cite {Zhang2012unconventional}, forming a single or a multiple of weak superconducting thin slices normal to the \textit{a} axis. As a result, vortices, as locally quenched superconductivity, will most probably reside inside the weak superconducting slices (more energetically favorable), and may enormously elongate along the slices \cite{Liu2011vortex, Song2012suppression}. However, our Bogoliubov de-Gennues (BdG) self-consistent calculations show that this explanation from the quantum confinement effect is applicable only if the Fermi wave length $\lambda_\textrm{F}$ is larger than about eight times of the lattice constant. In a usual metal, however $\lambda_\textrm{F}$ is commonly comparable to its lattice constant (e.g.\ $\lambda_\textrm{F}\sim 0.4 a_0$ in $\gamma$-Ga) \cite{Zhang2015molecular}, as indicated by considering free electron approximation. Therefore, the quantum confinement effect might not play the primary role in the vortex elongation observed here.

\begin{figure}[t]
\includegraphics[width=\columnwidth]{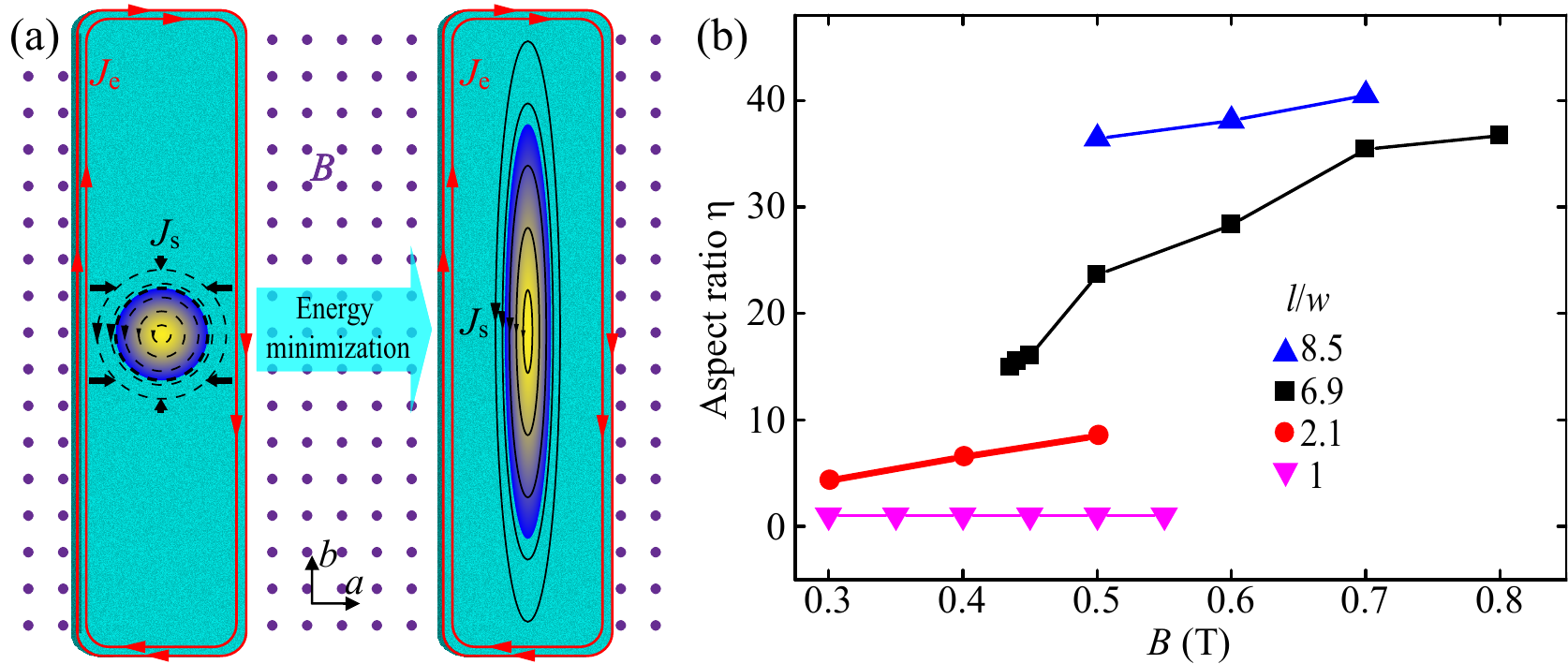}
\caption{(color online) (a) Schematic of vortex axial elongation. $J_e$ and $J_s$ denote the Meissner screening currents at the sample edge and around the vortex, respectively. Black thick arrows label the repulsive force between the vortex and Meissner screening currents $J_e$. (b) Magnetic field and $\gamma$-Ga geometry dependence of vortex elongation $\eta$. The extracted $\eta$ for $l/w$ = 1.0, 2.1, 6.9 and 8.5 are coded magenta, red, black and blue, respectively.
}
\end{figure}

Alternatively, another possible origin of elongated vortices is from the geometric confinement effect imposed by narrowness of the rectangular Ga nanostrips. In the Ga nanostrips studied here whose widths are of the order of $\xi$ (e.g.\ $w \sim 2.5\xi$ in Fig.\ 2), the finite size of vortex core cannot be ignored. Moreover, the nanostrip thickness (several nanometers) is considerably smaller than the penetration depth $\lambda$, vortices are of the Pearl rather than Abrikosov type \cite{Pearl1964current}. The vortex interactions with the Meissner screening currents at the sample edge ($J_e$) becomes long ranged and prominent. In terms of the counterpropagation between $J_e$ and the screening current $J_s$ encircling the vortex (left panel in Fig.\ 4(a)), the interactions are dominantly repulsive and pronounced along the \textit{a} axis. To minimize this repulsive interaction, the vortex current $J_s$ will deform or more specifically elongate along the \textit{b} axis until they are fully exploring the sample geometry (right panel in Fig.\ 4(a), less energetically costly), where the $J_e$-$J_s$ repulsive interaction along the \textit{b} axis gets stronger as well. Consequently, an equilibrium with the strong deformation in $J_s$ and thus ZBC or vortex cores is reached, as observed.

From considering $J_e$-$J_s$ repulsive interactions, one may further argue that vortex elongation $\eta$ could enhance as the ratio $l/w$ and magnetic field increases, because $J_e$ increases with increasing field. In addition, once if multiple vortices enter into the nanostrips, the strong interaction between vortices will not only compress a certain vortex along the \textit{a} axis as well and contributes to enhance $\eta$, but also configure all vortices parallel to the \textit{b} axis, as observed. Figure 4(b) summarizes the vortex elongation $\eta$ as function of the field \textit{B} and $l/w$ of the $\gamma$-Ga nanostructrues. Clearly, $\eta$ increases abruptly with \textit{B} and $l/w$. For example, in the Ga nanostrip with the aspect ratio $l/w$ of 8.5, $\eta$ is as high as 40.5 at 0.7 T. Therefore, even though further detailed theoretical investigations will be needed to understand the vortex elongation in a quantitative way, our experimental findings support that the repulsive interaction between the vortices and sample edge screening current is a reasonable explanation of the vortex elongation in Ga nanostrips.

To summarize, our real space visualization of unusually elongated vortices in $\gamma$-Ga nanostrips sheds light on Pearl vortex dynamics in the extremely thin and narrow superconducting nanostrips ($w <4\xi$), where the geometrical confinement effect may deform the internal vortex core structure and then configure vortices in a brand-new way. The elongated vortices observed here provide a novel platform to study exotic vortices in nanoscale superconductor. From the viewpoint of superconductor applications, the strong geometrical confinement could impose deep potential well for the vortex transverse motion and immobilize the vortices, providing a promising strategy to design nanoscale superconducting devices working at high magnetic field, a long-standing dream of superconducting research.

\begin{acknowledgments}
This work was supported by National Science Foundation and Ministry of Science and Technology of China. 
\end{acknowledgments}

%\bibliography{Gafilms}
%

\section{Supplemental Material}

\setcounter{figure}{0}
\setcounter{equation}{0}
\makeatletter
\renewcommand{\thefigure}{S\@arabic\c@figure}
\renewcommand{\theequation}{S\@arabic\c@equation}
%\maketitle

Commercially purchased Si(111) wafers with a resistivity of 0.01 $\Omega\cdot$¡¤cm were used as substrate, and the clean Si(111)-7 $\times$ 7 surfaces were obtained by flash annealing to $\sim$ 1200$^\textrm{o}$C while keeping the vacuum better than 1 $\times$ 10$^{-9}$ Torr. We then prepared Ga nanoislands by evaporating Ga (99.999$\%$) sources from a standard Knudsen cell at a nominal beam flux of approximately 0.4 monolayer (ML, $\sim$ 0.13 nm)/min, during which the Si(111)-7 $\times$ 7 substrate was kept at 110 K. Subsequent post-annealing at a slightly higher temperature of 150 K led to many Ga nanostructures with varying geometries and the lateral dimensions \cite{Zhang2015molecular}. Analogous to MBE growth of Pb on Si(111)-7 $\times$ 7 \cite{jia2007quantum}, the substrate just beneath Ga islands consists of a slightly disordered Ga wetting layer (1$\sim$2 ML), which exhibits no superconductivity down to the minimum measurement temperature of 2.2 K.

\begin{figure}[h]
\includegraphics[width=0.68\columnwidth]{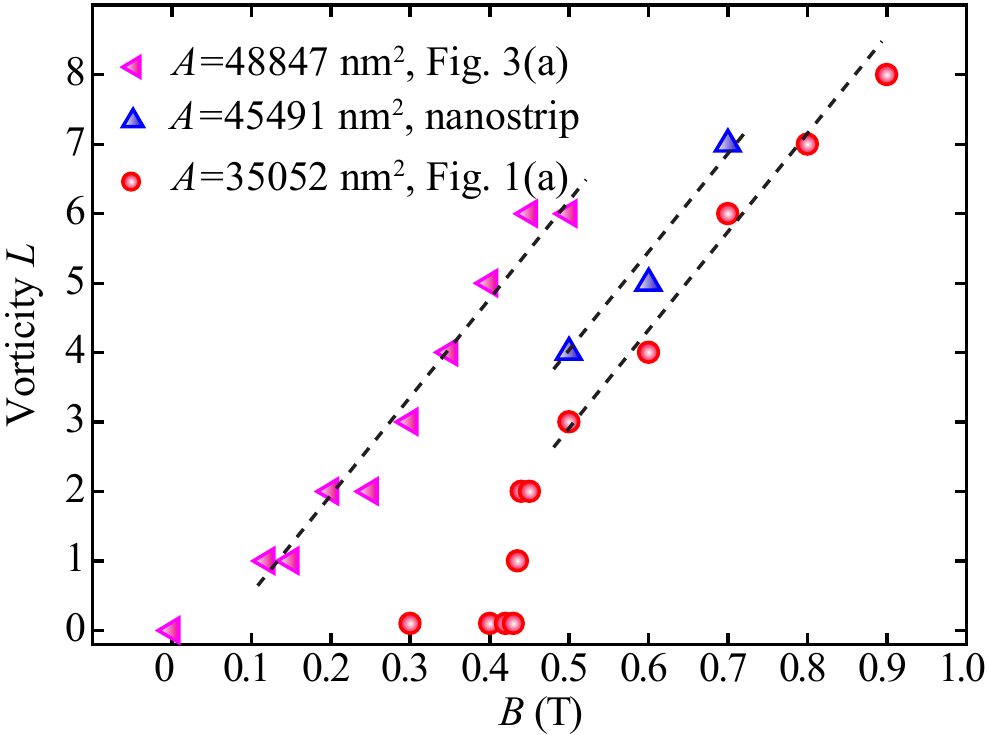}
\caption{Vorticity $L$ plotted as a function of the applied field $B$ for three Ga nanaoislands with varying geometry and lateral area $A$.
}
\end{figure}
Figure S1 depicts the evolution of vorticity $L$ (the number of individual isolated stripy or round yellow features in ZBC maps) with varying magnetic field $B$. Above a threshold field $B_\textrm{m}$, depending on $A$ or the width $W$ of superconducting strips, $L$ increases almost linearly with $B$, as anticipated for the gradual penetration of magnetic vortices into Ga islands \cite{Stan2004critical}. By extracting the slope (dashed lines) and island area $A$, we estimate that each of such features accommodate the magnetic flux of 1.18$\Phi_0$ (left triangle), 1.25$\Phi_0$ (upper triangle) and 1.47$\Phi_0$ (circle), respectively, close to a single magnetic flux quantum $\Phi_0$ and never exceeding 2$\Phi_0$. This strongly supports a single magnetic vortex nature of the observed stripy or round features in Figs.\ 2 and 3. The small discrepancy from $\Phi_0$ most possibly originates from an overestimate of the effective island area $A$ for nanoscale superconductors. Compared to Nb strips previously reported \cite{Stan2004critical}, $B_\textrm{m}$ appears larger in Ga strips we studied here, due to their extremely small dimensions.

\begin{figure}[h]
\includegraphics[width=0.9\columnwidth]{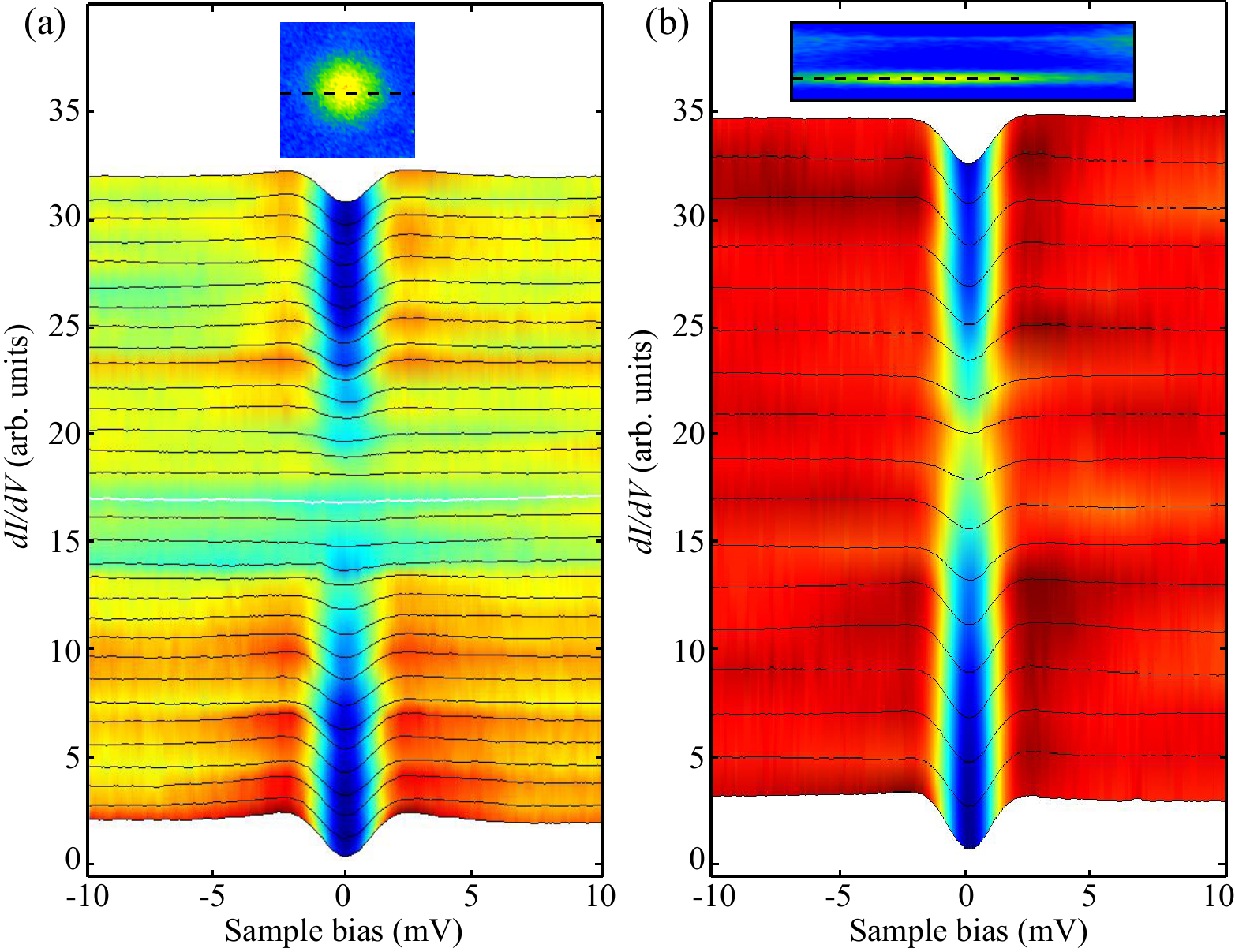}
\caption{Differential conductance $dI/dV$ spectra straddling (a) a round magnetic vortex core (Fig.\ 3, 0.15 T) and (b) an elongated vortex core (Fig.\ 2, 0.435 T), respectively. Inserted are the two vortices, with the dashed lines indicting where the $dI/dV$ spectra are acquired. As anticipated, the superconductivity is significantly suppressed with enhanced ZBC when approaching the vortex cores, regardless of their internal core structure.The small $E_F$-near conductance reduction in (b) is primarily caused by a large ($\sim$ 12 nm) spatial interval between the neighboring $dI/dV$ spectra, which fails to acquire the spectrum justly at the vortex core center.
}
\end{figure}

%\bibliography{Gafilms}

\begin{thebibliography}{32}
\bibitem [{sup()}]{supplementary}%
  \BibitemOpen
  \href@noop {} {\bibinfo  {journal} {Materials and Method are available as
  supporting material}\ }\BibitemShut {NoStop}%
\end{thebibliography}
%

\end{document}